\documentclass[aip, print]{revtex4-1}

\usepackage{graphicx}
\usepackage{dcolumn}
\usepackage{bm}
\usepackage{sidecap}
\usepackage{xcolor}
\draft 

\begin{document}

\title{Hyper Resolution Two Photon Direct Laser Writing using ENZ Nano-Cavity} 

\author{Giuseppe Emanuele Lio}
\affiliation{Physics Department, University of Calabria, I-87036 Arcavacata di Rende (CS), Italy}
\affiliation{CNR Nanotec-Institute of Nanotechnology, UOS Cosenza, 87036 Rende (CS), 87036, Italy}
\author{Tiziana Ritacco}
\affiliation{CNR Nanotec-Institute of Nanotechnology, UOS Cosenza, 87036 Rende (CS), 87036, Italy}
\affiliation{Physics Department, University of Calabria, I-87036 Arcavacata di Rende (CS), Italy}
\author{Antonio Ferraro}
\email[]{E-mail: antonio.ferraro@unical.it}
\affiliation{Physics Department, University of Calabria, I-87036 Arcavacata di Rende (CS), Italy}
\affiliation{CNR Nanotec-Institute of Nanotechnology, UOS Cosenza, 87036 Rende (CS), 87036, Italy}
\author{Antonio De Luca}
\affiliation{Physics Department, University of Calabria, I-87036 Arcavacata di Rende (CS), Italy}
\affiliation{CNR Nanotec-Institute of Nanotechnology, UOS Cosenza, 87036 Rende (CS), 87036, Italy}
\author{Roberto Caputo}
\email[]{E-mail: roberto.caputo@unical.it}
\affiliation{Physics Department, University of Calabria, I-87036 Arcavacata di Rende (CS), Italy}
\affiliation{CNR Nanotec-Institute of Nanotechnology, UOS Cosenza, 87036 Rende (CS), 87036, Italy}
\affiliation{Institute of Fundamental and Frontier Sciences, University of Electronic Science and Technology of China, Chengdu 610054}
\author{Michele Giocondo}
\affiliation{CNR Nanotec-Institute of Nanotechnology, UOS Cosenza, 87036 Rende (CS), 87036, Italy}

\begin{abstract}
A novel technique is reported to improve the resolution of two-photon direct laser writing lithography. Thanks to the high collimation enabled by extraordinary  $\varepsilon_{NZ}$ (near-zero) metamaterial features, ultra-thin dielectric hyper resolute nanostructures are within reach. With respect to the standard direct laser writing approach, a size reduction of $89\%$ and $50\%$ , in height and width respectively, is achieved with the height of the structures adjustable between 5nm and 50nm. The retrieved 2D fabrication parameters are exploited for fabricating hyper resolute 3D structures. In particular, a highly detailed dielectric bas-relief (500 nm of full height) of Da Vinci's \textit{"Lady with an Ermine"} has been realized. The proof-of-concept result shows intriguing cues for the current and trendsetting research scenario in anti-counterfeiting applications, flat optics and photonics.
\end{abstract}

\pacs{}

\maketitle 
\section{Introduction}
A variety of wavelengths of the electromagnetic spectrum are typically involved in the fabrication of structures with details ranging from micro- to nano-scale exploiting a plethora of techniques including holography, laser ablation and  UV lithography \cite{infusino2012polycryps, sahin2014nanoscale, sze1985physics}. Large area ($cm^2$) and features of hundreds of nanometers are at hand in systems like microfluidic channels, optical devices, MEMs, and transistors, just to name a few \cite{leclerc2004microfluidic, psaltis2006developing}. Pushing the resolution to nanometric scale requires a different approach like nanoimprinting and electron beam lithography, albeit at the expense of reduced work area and time-consuming procedure \cite{ferraro2018directional, sze1985physics}. 
In the last decades, the new frontier of nanometric fabrication is embodied by two photon direct laser writing (TP-DLW) lithography. By exploiting a nonlinear two-photon absorption process, the involved photoresin is cured only in the focal point of the used laser, the voxel (short for volume pixel), thus sensibly increasing the resolution of realized nanostructures. However, nanotechnology still moves forward seeking hyper-resolution and recent years have witnessed a very large number of attempts  to further increase the fabrication performance. Among them, chemical reagents have been involved to improve the photoresin capabilities \cite{zhang2010designable}, or complex reaction procedures, as modification of initiation and termination polymerization phases by means of a gain medium have been considered. In the latter case, a resolution of $\sim \lambda/20$ has been achieved \cite{li2009achieving}. 
Results are also achieved in the TP-DLW technique in terms of 3D micro scale structures \cite{zhang2010designable}, for improving as example fiber tip fabrication \cite{bratton2006recent} and high resolution of 3D systems in hydrogels \cite{xing20153d}. 
In the last decade, in terms of achieved resolution, a $100nm$ limit was possible with the use of radical quenchers in femtosecond laser direct writing \cite{lee2008advances}, of $40nm$ by using an activation beam \cite{li2009achieving}, of $25nm$ and $20nm$ using the scan speed manipulation and self-smoothing effect \cite{tan2007reduction, wu2010high}, or using  photosensitive sol-gels \cite{passinger2007direct}. Finally, a technical expedient, like changing the interface height, can limit the single line width and height to around $100nm$ \cite{park2009two}.
A completely different approach involves  optical epsilon-near-zero ($\varepsilon_{NZ}$) nano-cavities in the metal/insulator/metal/insulator (MIMI) configuration \cite{liocolor}. These systems pave the way for the realization of very versatile devices with unusual optical features. In this manuscript, the extraordinary self-collimation of light enabled by a MIMI plasmonic metamaterial is proposed as a ground-breaking possibility to improve the resolution of TP-DLW lithography. In particular, nanostructures with typical sizes of few tens of nanometers are within reach in few minutes writing time. 
In the following, it is shown how the metamaterial is exploited to successfully fabricate 1D gratings with a height adjustable from $5 nm$ to $50 nm$ and a nanometric 3D bas-relief of Da Vinci's portrait \textit{"Lady with an Ermine"} with the dimension of $50X80X0.5 \mu m$ with its full height divided in $25$ slices of $20nm$ thickness each. The proposed approach is characterized by a very fast patterning process, able to realize dielectric nanometric structures with almost any shape/form as ultra-thin diffractive optical elements \cite{chen2013reversible, chen2018broadband}. 
Moreover, the proposed MIMI can be also realized directly over the objective of the TP-DLW apparatus enabling the nanafabrication on different substrates. 
These results are extremely important for industrial applications in several fields such as anti-counterfeiting, flat optics.

\section{Principle and Materials of Hyper Resolute Two Photon Lithography}
\begin{figure}[ht] 
\begin{center}
\includegraphics[scale=0.2]{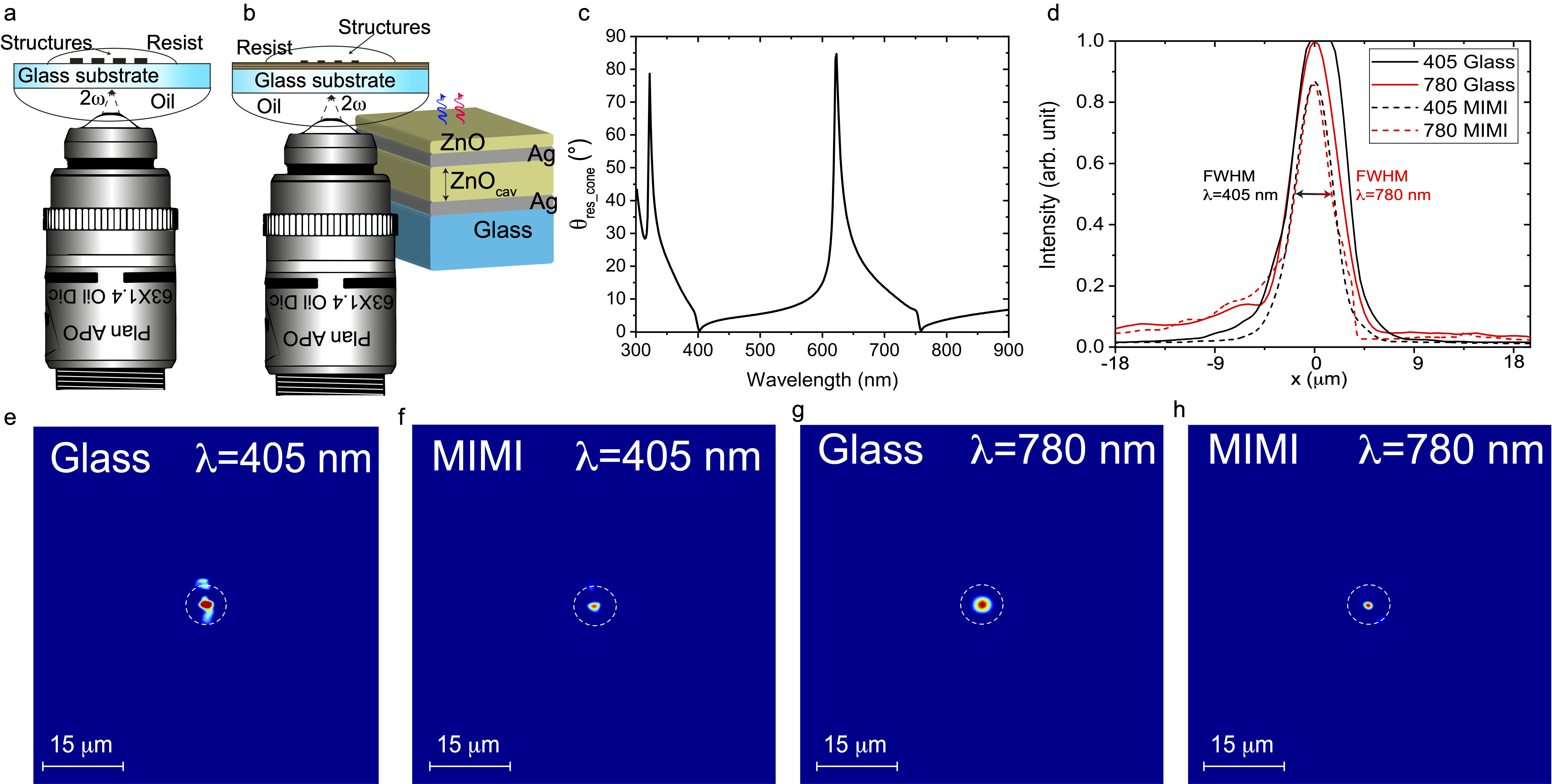}
\caption{Sketch of TP-DLW process where the photo-sensible resin is placed a) on the glass substrate, b) on the MIMI device working as substrate. The  inset show the proposed MIMI structure made by Ag/ZnO/Ag/ZnO allowing the transmission of wavelengths of 390 and 780 nm. c) Resonant cone angle for the considered MIMI. d) Point spread function (PSF) for the four considered cases . The PSF reveals a beam shrinking about $36 \%$ with MIMI respect to the glass. e-f) Transmitted spot trough glass and mimi at $\lambda 405 nm$ and e-f) at $\lambda=780 nm$, acquired trough beam profiler.}
\label{1}
\end{center}
\end{figure}
The TP-DLW apparatus used in this work exploits a femtosecond Ti:Sapphire laser ($\lambda = 780 nm$) connected to an inverted microscope. The laser beam is focused on the sample through a $63X$ objective, with 1.4 numerical aperture (NA). 2D and 3D structures can be fabricated through the glass by diving the objective in the photoresist (Fig. \ref{1}a). 
Since the writing process occurs plane by plane, to create the highest reliefs of a structures, the laser beam is attenuated during the travel through the objects already fabricated, generating smaller effective voxels.
As illustrated in Fig. \ref{1}b, it is possible to drastically increase the resolution in terms of voxel sizes by using a MIMI directly deposited on a classical substrate used for the TP-DLW. The inset reports a schematic view of the MIMI constituted of a silver layer with thickness $t_{Ag}=30nm$,  a thicker ZnO layer working as an optical nano-cavity ($t_{cav}$), another silver layer with the same ($t_{Ag}$) thickness and a final thin ZnO layer ($t_{ZnO}=30nm$). 
In order to obtain a hyper-resolute TP-DLW process, it is necessary that selected wavelengths ($\lambda=780nm$ and $\lambda=390nm$) are let through the MIMI device. By using a Finite Element Method (FEM) model, based on numerical ellipsometer analysis (NEA)\cite{lio2019comprehensive}, and by varying the thickness of the dielectric nano-cavity ($t_{cav}$), it is possible to retrieve the thickness value to obtain the minimum in reflectance and the maximum in transmittance for a normally incident wave at the two above mentioned wavelengths. In our case, the optimal cavity thickness is $160 nm$ that actually supports the plasmonic resonant modes marked with dashed white lines in the reflectance and transmittance maps reported in Figure S1a and b, respectively.  
A further validation of the occurrence of double modes is provided by a modified effective medium theory (EMT), taking into account the dielectric constant and the thickness of each layer  \cite{zeng1988effective, rousselle1993effective}. This analysis shows the double $\varepsilon$NZ behavior of the proposed MIMI device presenting the zero crossing point between the real perpendicular $Re \left\{ \varepsilon_{\perp}\right \}$ and parallel $Re \left\{ \varepsilon_{\parallel}\right \}$ dielectric constant respectively for the two resonant wavelengths (Figure S1c in supplementary material).
In case of off-normal light incidence on the same MIMI ($\theta_i$ varying from $0^\circ$ to $80^\circ$) reflectance and transmittance curves are reported in Figures S1d-g. Experimental curves are measured by analyzing the fabricated MIMI by means of  a W-VASE ellipsometer, while the numerical analysis is performed again by the NEA model. The quite reliable agreement between numerical and experimental results confirms the possibility to exploit the MIMI for TP-DLW with a focused ($\lambda = 780 nm$) laser beam at normal incidence. As reported in past studies \cite{mocella2009self, polles2011self, di2012digital, arlandis2012zero}, $\varepsilon_{NZ}$ metamaterials have the remarkable ability to collimate light.

The propagation of light within these metamaterials can be rigorously described by means of the dyadic Green's function \cite{potemkin2012green}. Such an analysis confirms that light emitted in the direction of the extraordinary axis from a localized source placed on the top of $\varepsilon_{NZ}$ metamaterials, propagates within the medium in the so-called resonance cone \cite{newman2013enhanced}. The resonance cone is visible as two lobes propagating through the medium, separated by a semi-angle $\theta_{res-cone}$, which is calculated as follows \cite{shekhar2014hyperbolic}: 
\begin{equation}
\left | \theta_{res-cone}  \right |= \sqrt{-\frac{\varepsilon_\parallel}{\varepsilon_\bot}} 
\end{equation}
For the considered MIMI device the $\theta_{res-cone}$ has been calculated using the parallel and perpendicular dielectric constant retrieved with the EMT. As reported in Figure \ref{1}c, the MIMI device presents two points where the resonant cone angle is zero meanimg that, for $\lambda\sim 400nm$ and $\lambda\sim770nm$, the light passing through the MIMI remains completely collimated. Due to the incident angle independency in the EMT and $\theta_{res-cone}$ calculation, the mismatch in terms of desired and evaluated wavelengths is not worrisome. 
The $\theta_{res-cone}$ and the beam behavior is experimentally confirmed by the use of a homebuilt confocal setup (details reported in Figure S2). Each spot has been collected using a beam profiler (Thorlabs BC106N-VIS spectral range from $350 nm$ to $1100 nm$).
The MIMI is compared with a standard glass (glass coverslip $24X24X0.15 mm$) used in TP-DLW, the produced focalized spot is measured in order to estimate the beam shrinking by the point spread function (PSF) as illustrated in Figure \ref{1}d.
The intensity maps are  reported in Figures \ref{1}e-h. The beam with wavelength of $\lambda=405 nm$ and $\lambda=780 nm$ passing through the MIMI presents a reduction of the initial full width at half maximum (FWHM) of $1\mu m$ and about $0.9 \mu m$, respectively, which corresponds to a reduction of $36 \%$ when collected with a 50x objective. In case of a 10x objective, the beam reduction that occurs is about $20\%$ as detailed in the field maps and PSF reported in Figure S3. 
\section{Sample Fabrication and Characterization} 
\begin{figure}[!ht]
\begin{center}
\includegraphics[scale=0.2]{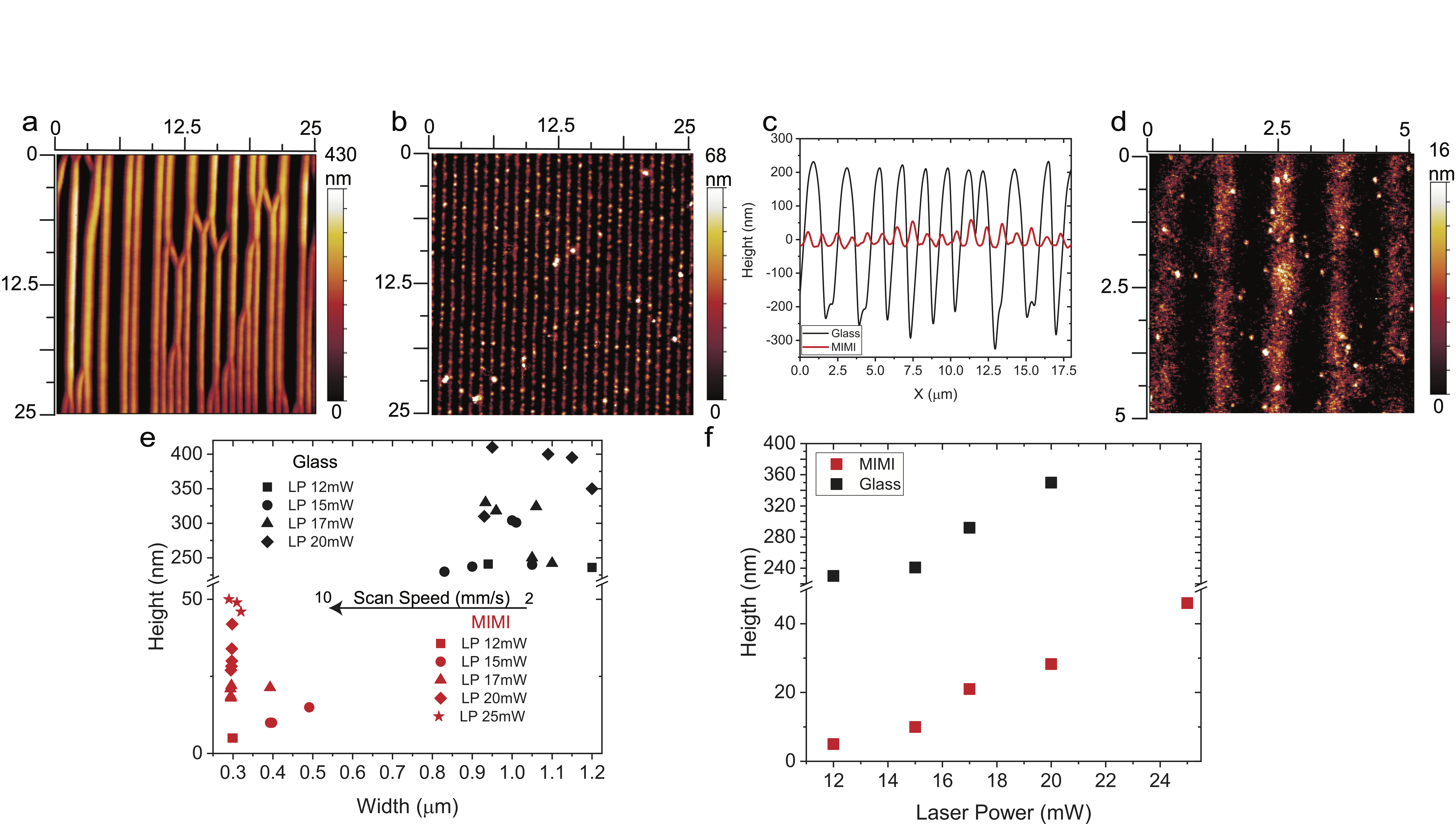}
 \caption{ AFM morphology of a fabricated grating  with a Laser Power (LP) of $20 mW$ and a Scan Speed (SS) of $4 mm$ a) trough the classical glass substrate and b) trough the MIMI device. c) Profiles comparison of the two fabricated grating. d) Grating realized trough the MIMI device using a LP of $12 mW$ and a SS of $4 mm$ producing an element height of $\sim 5\pm2 nm$. e) Height and Width for all the fabricated gratings at different laser power and scan speed using the glass substrate (black symbols) and the MIMI (red symbols). f) Height as function of the LP for glass (black square) and MIMI (red squares) substrates showing a reduction of $\sim89\%$.  }
\label{2}
\end{center}
\end{figure}
\begin{figure}[ht]
\begin{center}
\includegraphics[scale=0.6]{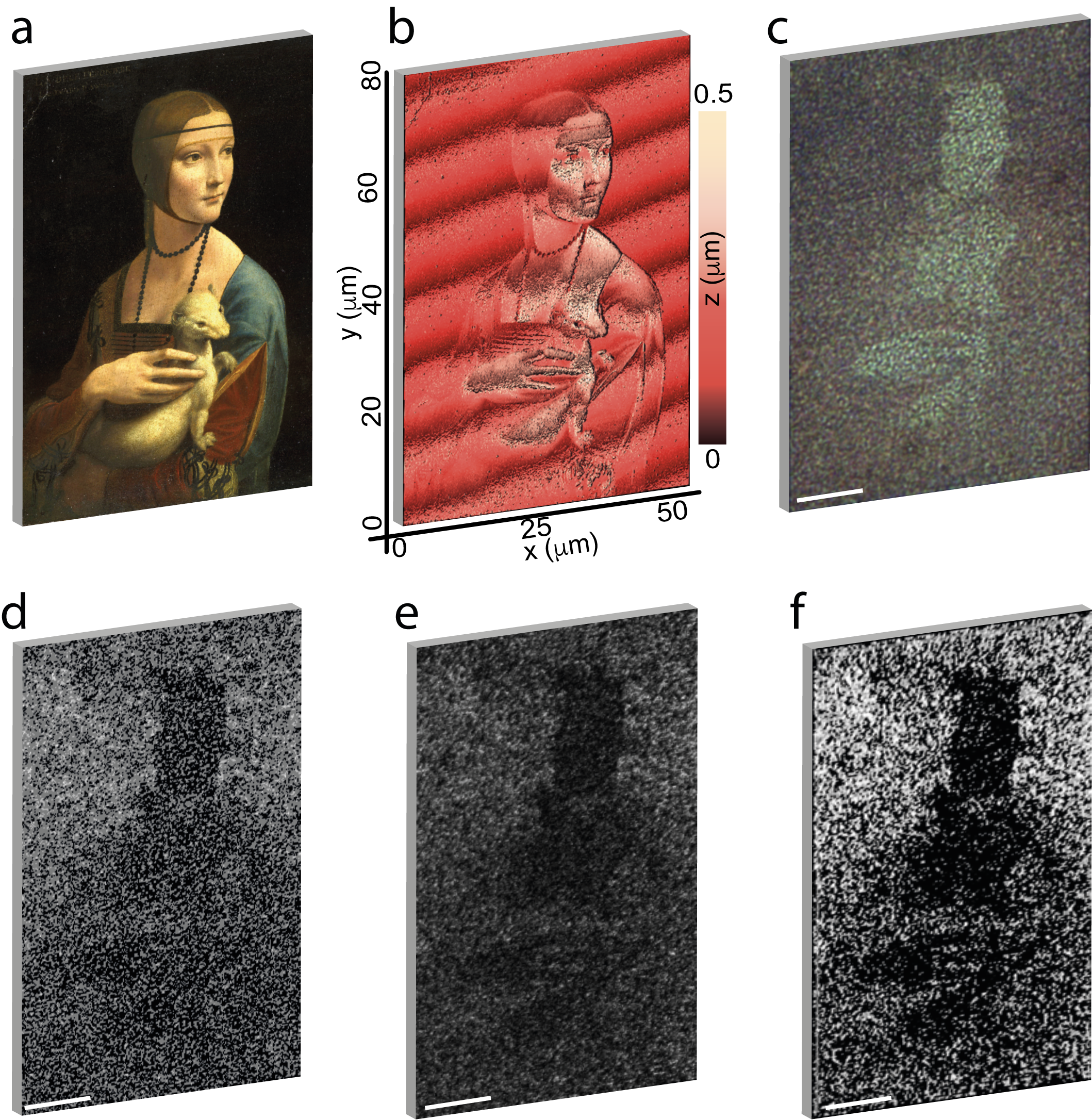}
 \caption{a) Original portrait and b) TP-DLW software model of the bas-relief \textit{"Lady with an Ermine"}. c) Optical image of the realized 3D sample collected with an objective with magnification 40X trough the glass and enlighten from the top. d-f) Emission confocal images of the bas-relief. In particular d) reports the base of the 3D print showing the silhouette of the "Lady". e) Image collected close to the top ($H_z\sim 0.5 \mu m$) of the portrait  and some details are shown as the hand, the shoulders, and  parts of dress and face. f) A whole sample image has been reconstructed highlighting the reliability to "print" very well all particulars at the nanoscale. The scale-bar in each picture is $15 \mu m$. }
\label{3}
\end{center}
\end{figure}

To demonstrate the hyper resolution in a TP-DLW process induced by the presence of a MIMI, an array of 1D gratings has been fabricated on top of the MIMI previously deposited on a glass coversplip substrate. In the latter case, the interface selected to begin the laser writing process is established between the resin and the glass; while in presence of the MIMI, the chosen interface is between the first silver layer and the glass. The use of an array permits to characterize each elements as a function of the laser power $LP$, which varies from $12 mW$ to $\sim25 mW$, and the scan speed $SS$ which varies from $2 mm/s$ to $10 mm/s$. The topography analysis conducted using an atomic force microscope (AFM) with high precision tips (see details in the methods section) ensures to collect very high quality data. A comparison between the AFM measurements performed on the structures realized through a simple glass coverslip and through the MIMI is shown in the Figure \ref{2}. The difference between the standard process and through the MIMI stands out immediately offering the possibility to clearly trace straight lines without defects (Figures \ref{2}a and b). A comparison between two fabricated gratings, using $LP = 20 mW$ and $SS = 4 mm/s$, has been done based on the raw images of the AFM profile. As illustrated in Figure \ref{2}c, a remarkable difference in terms of height and width of each grating element is present, with  a height $H\sim 420nm$ for the grating made trough glass and an average $\tilde{H}\sim{30}nm$ and and same value for the half width of the MIMI.
In Figure \ref{2}d, a grating image is reported realized using the lowest laser power and the fastest scan speed producing an $\tilde{H}\sim{7}nm$ and a width $\tilde{W}\sim{150}nm$. In order to complete and  better understand the comparison on both substrates, the size of each grating element contained in the test array have been collected and reported in the graph shown in Figure \ref{2}e.  The latter evidences  an interesting and remarkable difference between the grating sizes produced through the glass and the MIMI. In fact, the resolution is improved in percentage by $\sim89\%$ and $\sim50\%$ in terms of height and width respectively.  A further validation has been done by comparing the produced element height at different laser powers. The trend is similar for both substrate maintaining equal the difference that occurs between them as presented in Figure \ref{2}f.
The proof of concept of the reliability of the proposed technique, able to force the beam self-collimation during writing, is the fabrication of particularly complex three-dimensional TP-DLW tests. The MIMI functionality has been challenged in producing a polymer bas-relief version of the famous Da Vinci's portrait \textit{"Lady with an Ermine"} (Figure \ref{3}a). The choice to realize a bas-relief has been done considering that the process involves a 3D lithography and this portrait also contains very tiny details only reproducible in presence of hyper resolution. The first step is create a computer-aided design (CAD) of the portrait image to be used in the TP-DLW process. The obtained design is shown in Figure \ref{3}b. The full height of the portrait is chosen as $H_z= 500nm$ that is divided in $25$ slices of $20nm$ thickness each. The optical microscopy image collected by using unpolarized white light and a $40x$ objective clearly underlines the contours of the portrait and details like face, dress, hand, ermine and necklace (Figure \ref{3}c). This high quality optical image is also the consequence of the self-collimation of the microscope impinging light passing through the sample and reaching the objective experimenting the lens effect of the $\epsilon_{NZ}$ metamaterial as demonstrated in previous works\cite{fang2002imaging, liu2007far, casse2010super, zhao2011nanoscale, kim2015metamaterials}. The collection of the whole bas-relief (Figure \ref{3}d-e) is obtained by means of a fluorescence confocal microscopy analysis (details in the methods section) performed with a slicing of $10 nm$ along the z-axis . This imaging method allows to easily recognize at the minimum height the silhouette of the "Lady" (Figure \ref{3}d). Then by increasing the $H_z\sim 0.5\mu m$, the frame shows the details present on top of the sample, as illustrated in Figure \ref{3}e. Finally, once the z-stack is completed, by using the proprietary software is possible to recombine all acquired images producing their full overlap and the final picture reported in Figure \ref{3}f. 
The preliminary test highlights the reliability of the hyper resolute 3D TP-DLW with slice distance of $20 nm$ and an high level of details reproduced in a very small volume. This technique is foreseen as a valuable possibility to realize anti-counterfeiting tags in the recent research framework of physical unclonable functions. 

\section{Conclusions}
In this work, a novel technique is presented to sensibly improve the performance, in terms of writing resolution, of a generic two-photon direct laser writing process. The diffraction-free ability of an $\varepsilon_{NZ}$ metamaterial enables an extraordinary collimation of the writing laser light and hence the hyper resolution of the TP-DLW. In fabricating test structures like 1D gratings, a reduction of the voxel size of about $89\%$ and $50\%$, for height and slit width respectively, is observed corresponding for the height of the grating to a reduction from $250\pm2 nm$ to $5\pm2 nm$.
The proposed technique gives its best when more complex 3D structures are considered. An hyper-resolute bas-relief version of the famous Da Vinci's portrait \textit{"Lady with an Ermine"}  with a full height of only $500nm$ divided in $25$ slices of $20nm$ thickness each. These new frontier results find immediate application in the trendsetting scenario of physical unclonable functions and flat-optics. 
\section*{Methods}
\textbf{Sample fabrication} The MIMI substrate is realized by DC sputtering deposition on glass substrate (thickness $\sim 150\mu m$), 
Then, the substrate is placed inside the two-photon lithography apparatus which is able to define the grating pattern or the desired shape in a drop of photo-resine placed on top of the MIMI device. Finally, the sample is developed in a bath of propylene glycol methyl ether acetate (PGMEA) for 25 minutes, then in a bath of isopropanol alcohol (IPA) for other 5 minutes. During all the development process, the sample is soaked.\\
\textbf{AFM characterization} has been done using a confocal microscope Zeiss LSM 780 equipped with an AFM head-stage. The AFM measurements are performed with high resolute tips with precision of $\pm 2 nm$. Each scan has been collected at high resolution of 1024x1024 px in order to reduce any background noise. \\
\textbf{Fluorescence Confocal} 3D image in fluorescence (see Figures \ref{3}) has been acquired using a confocal microscope equipped with a 3D piezoelectric scanner. A 488 nm laser was focused on the cover-slip through a 40x air objective, which allows a spatial resolution of 200 nm, and a z-resolution of 10 nm. The emitted light collected by the objective is sent to a beam-splitter and filtered with a MBS T80/R20. Finally it passes through a pin-hole and is collected by a 580-600 nm detector. 
\section*{Contributions}
G.E.L. conceived the idea. G.E.L., T.R. and A.F. fabricated and fully characterized the structures. G.E.L. and A.F. performed the measurements and analyzed the data. R.C, A.DL. and M.G. devised the experiments and supervised the work. G.E.L. and A.F prepared and wrote the manuscript with input from all authors.
\section*{Acknowledgements}
The authors thank the Infrastructure ``BeyondNano" (PONa3-00362) of CNR-Nanotec for the access to research instruments. They also thank 
the ``Area della Ricerca di Roma 2", Tor Vergata, for the access to the ICT Services (ARToV-CNR). 
\bibliography{bibliography} 

\setcounter{figure}{0}
\renewcommand{\thefigure}{S\arabic{figure}}
\section*{Supplementary material}
\begin{figure}[ht]
\begin{center}
\includegraphics[width=0.8\columnwidth]{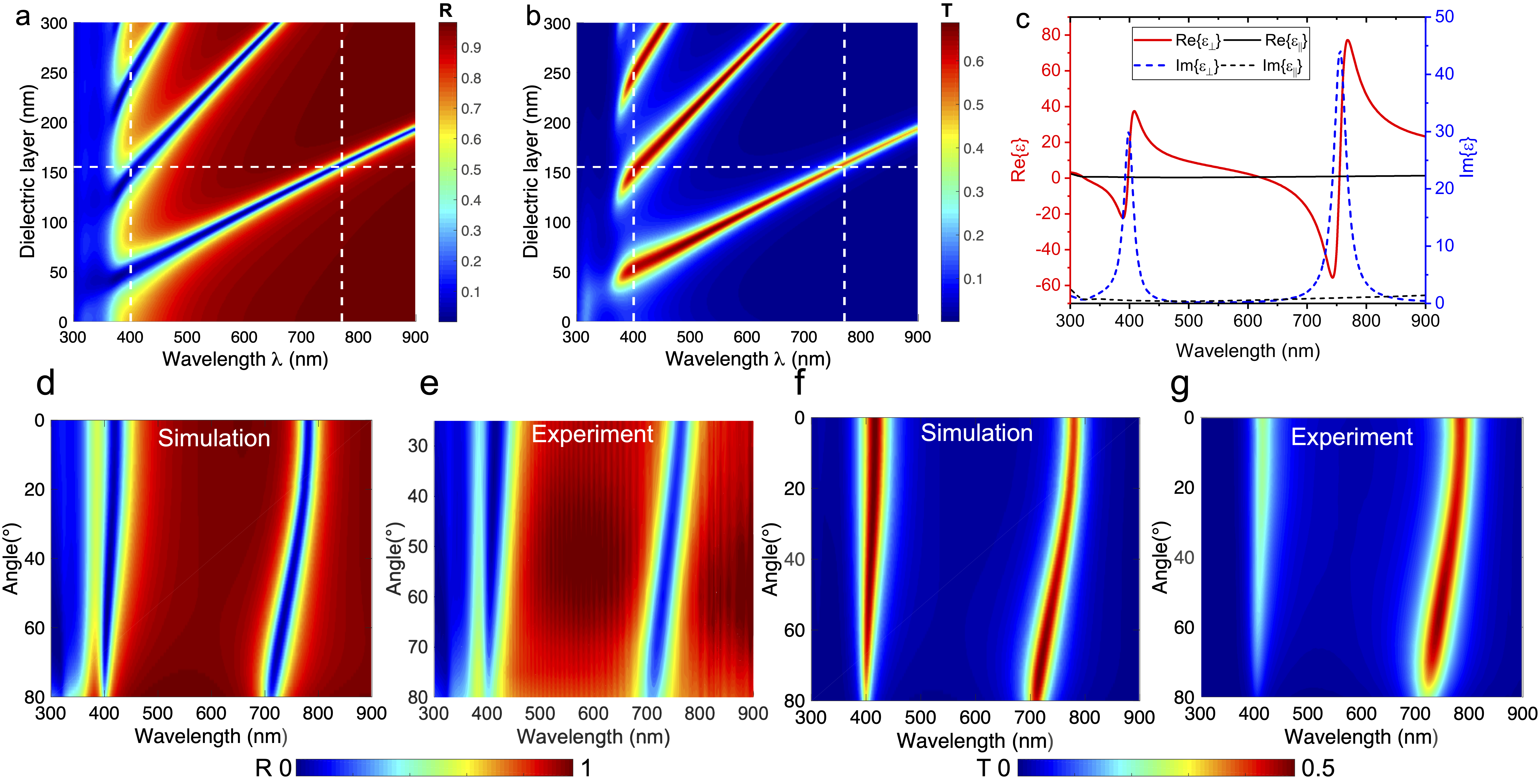}
\caption{a-b) Reflectance and transmittance map, respectively, at normal incidence varying the ZnO thickness that constitutes the optical nano-cavity. The value of $t_{cav}=160nm$ has been choose to obtain a minimum in reflectance and maximum in transmission at $\lambda=390nm$ and $\lambda=780nm$. c) Effective medium theory calculated for a four layer system showing the $\varepsilon NZ$ modes at the two resonant wavelengths. d-e) Numerical and experimental reflectance map and f-g) numerical and experimental transmittance map at different incident angle for the MIMI device showing the two minimum and two maximum respectively at $\lambda=390nm$ and $\lambda=780nm$.}
\label{S1}
\end{center}
\end{figure}
\subsection{Optical Setup}{
The optical setup for the Point Spread Function (PSF) characterization is composed as following. A collimation line for the white light that is produced by from a Xenon lamp is realized with a 40X objective and a fiber coupler mounted on a 3-axis stage. The collimated in sent to the main confocal line. On the same optical path, it has been placed another 3-axis stage with a 10x objective used only to detect the metalenses on the sample, in fact during the measurements it is moved backward in order to maintain only the collimated light that impinges on the metalens sample. Then, a 32x objective collects the signal from the sample and collimates it into the detector that is represented by a beam profiler (BP) Thorlabs BC106N-VIS spectral range from $350 nm$ to $1100 nm$ mounted on a double-axis stage. A lens collects the signal from the objective and sent it on the BP ccd. In the middle of this path there is the sample mounted on a holder stage that allows controlling micrometer movement. The sample is moved forward and backward along Z in order to collect the beam divergence. Between the fiber collimator and the first objective it has been placed a beam splitter that sends the light from lamp or from the lasers in the main path. The laser line is composed by different lasers. For example a blue laser impinges on a dichroic mirror (DM1) that reflects the green and leads both lasers on a beam splitter (BS1) that collects the red laser and sent all lasers on the second beam splitter (BS2). In order to measure the input power a flip mirror is placed before the 10X objective and the signal is sent on a Thorlabs power meter head (model S130VC). The whole described setup is depicted in Figure \ref{S2} 
\begin{figure}[ht]
\begin{center}
\includegraphics[width=0.8\columnwidth]{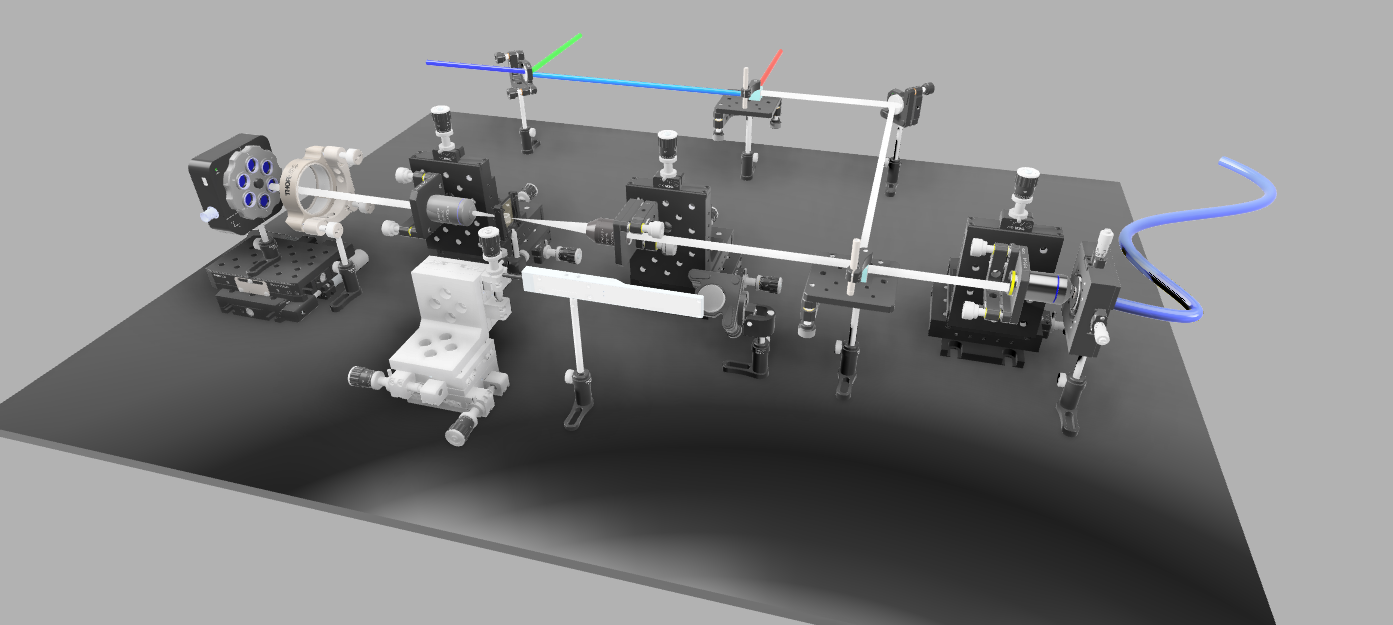}
 \caption{Schematic view of the optical setup used to fully characterize the produced PSF trough the MIMI device and the glass substrate.}
\label{S2}
\end{center}
\end{figure}
\begin{figure}[ht]
\begin{center}
\includegraphics[width=0.8\columnwidth]{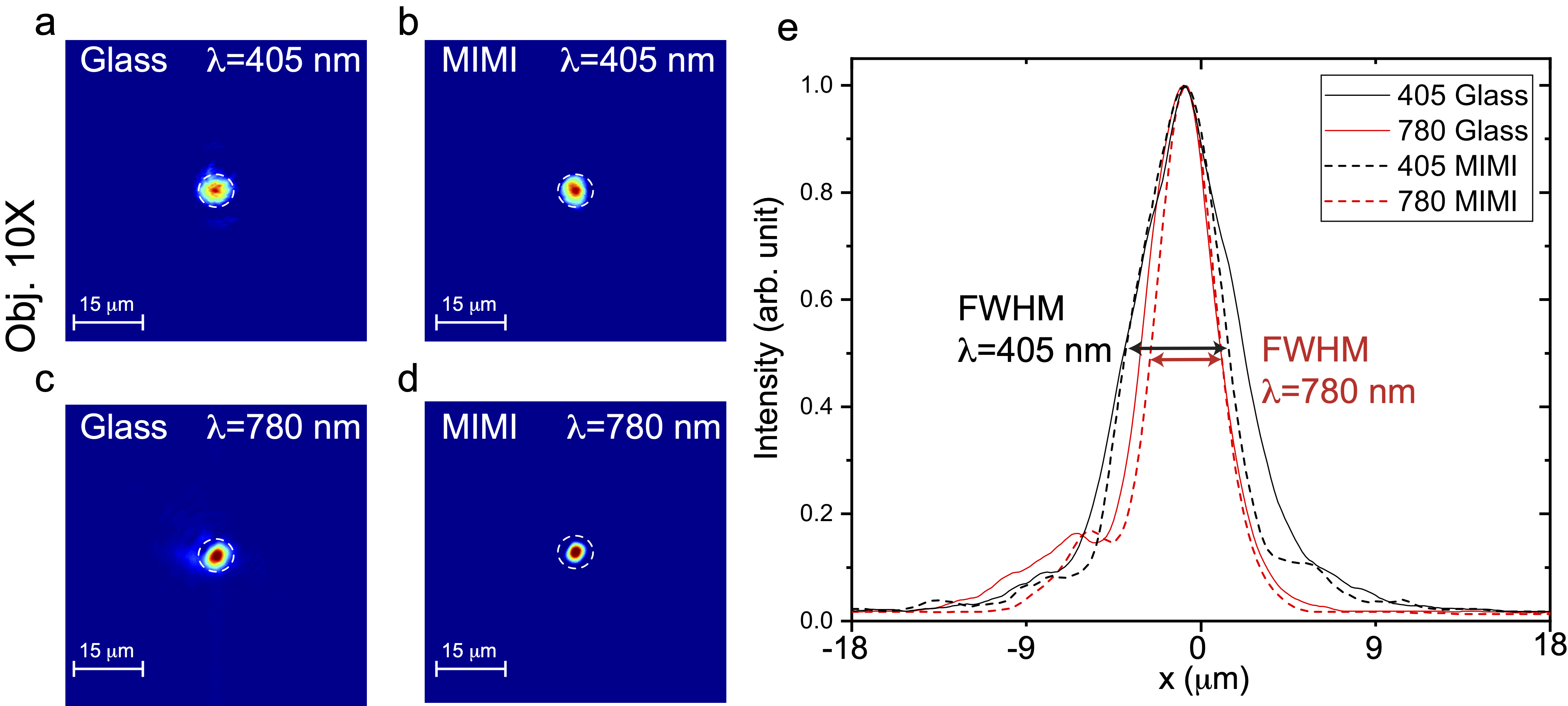}
 \caption{ a-b) Transmitted spot trough glass and MIMI at $\lambda 405 nm$ and c-d) at $\lambda=780 nm$, acquired trough the beam profiler. e) Point spread function (PSF) for the four considered cases . The PSF reveals a beam shrinking about $20 \%$ with MIMI respect to the glass.}
\label{S3}
\end{center}
\end{figure}
}
\end{document}